\documentstyle[prb,aps,preprint]{revtex}

\begin{document}

\draft 

\title{Current-voltage scaling of a Josephson-junction array at  
irrational frustration}

\author{Enzo Granato}
\address {Condensed Matter Physics Group, \\
International Centre for Theoretical Physics, \\
34100 Trieste, Italy \\
and \\
Laborat\'orio Associado de Sensores e Materiais, \\
Instituto Nacional de Pesquisas Espaciais, \\
12225 - S\~ao Jos\'e dos Campos, SP, Brasil \cite{byline}}

\maketitle

\begin{abstract}
Numerical simulations of the current-voltage characteristics of an
ordered two-dimensional Josephson junction array at an irrational flux
quantum per plaquette are presented. The results are consistent with an
scaling analysis which assumes a zero temperature vortex glass
transition. The thermal correlation length exponent characterizing this
transition is found to be significantly different from the
corresponding value for vortex-glass models in disordered
two-dimensional superconductors. This leads to a current scale where
nonlinearities appear in the current-voltage characteristics decreasing
with temperature $T$ roughly as $T^2$ in contrast with the $T^3$
behavior expected for disordered models.
  
\end{abstract}


There has been an increasing interest in two-dimensional
Josephson-junction arrays  as a model system to study disorder and
frustration effects as found in high-$T_c$ superconductors and
spin-glass systems \cite{physica}. In Josephson-junction arrays,
frustration without disorder can in principle be introduced  by
applying an external magnetic field on a perfect periodic array.  The
frustration parameter $f$, the number of flux quantum per plaquette, is
given by $f=\phi/\phi_o$, the ratio of the magnetic flux through  a
plaquette $\phi$ to the superconducting flux quantum $\phi_o = hc/2e$,
and can be tuned by varying the strength of the external field.
Frustration can be viewed as resulting from a competition between the
underlying periodic pinning potential of the array and the periodicity
of the vortex lattice \cite{teitel}. At a rational value of $f$, the
ground state is a commensurate pinned vortex lattice leading to
discrete symmetries in addition to the continuous $U(1)$
symmetry of the superconducting order parameter.  In particular, for
$f=1/2$ which has been intensively studied both experimentally and
theoretically, a superconducting phase transition takes place at finite
temperatures with an interplay of $U(1)$ and discrete $Z_2$ symmetry
\cite{granato}. At irrational values of $f$, the behavior is much less
understood since the vortex lattice is now incommensurate with the
array. The ground state consists of a disordered vortex pattern
\cite{halsey} lacking long range order which can also be regarded as a
vortex-glass state without disorder.  One can not completely rule out a
possible glass transition at finite temperatures as has been suggested
by Halsey \cite{halsey} but other arguments suggest a zero temperature
transition \cite{teitel,choi}.  In any case, on the basis of a close
analogy between this system and gauge glass models of disordered
superconductors, one expects metastable states with long relaxation
times and a nonzero critical current at zero temperature
\cite{halsey,choi}. In fact, measurements of I-V characteristics in
two-dimensional superconducting wire networks at an irrational
frustration \cite{goldman} have been interpreted within the scaling
analysis of a vortex glass transition as in three-dimensional
disordered superconductors where there is evidence of a
finite-temperature glass transition \cite{fisher1}. In the case of
superconducting wires, the behavior is likely to be dominated by a mean
field transition which should correspond to the behavior at higher
dimensions. This could provide a possible explanation for the observed
scaling behavior of a finite-temperature transition.  On the other
hand, no evidence was found in experiments on two-dimensional
proximity-coupled Josephson-junction arrays \cite{carini} where phase
fluctuations are expected to be more important. In disordered
superconductors, studies of different models of the vortex glass  tend
to agree that in two dimensions a vortex glass transition takes place
only at zero temperature \cite{fisher1} which is supported both by
numerical simulations \cite{fisher2} and experiments \cite{koch}.
Although being completely different in the nature of their ground
states, the similarity of the behavior of these two systems, specially
regarding slow relaxation vortex dynamics and a possible zero
temperature transition in low dimensions, strongly suggests that a
two-dimensional array at irrational $f$ should behave as a
zero-temperature vortex glass. In this case, the appropriate scaling of
the I-V characteristics should be the one corresponding to a zero
temperature transition while the different nature of the ground states
should be reflected in the value of the critical exponents as a
different universality class.  It  seems therefore worthwhile to
investigate to which extent an array at irrational $f$ can be described
as a zero temperature vortex glass.

In this work, we present simulations of the current-voltage
characteristics of a Josephson junction array at an irrational flux
quantum per plaquette $f=(3-\sqrt{5})/2$, a golden irrational, and an
scaling analysis \cite{fisher1} which assumes a zero temperature vortex
glass transition. The results are consistent with the scaling
assumption and allow for an estimation of the thermal correlation
length critical exponent $\nu $ characterizing the zero temperature
transition. This critical exponent is found to be significantly
different from the corresponding value for  vortex-glass models in
disordered two-dimensional superconductors.  As a result, the current
density scale, $J_{nl} \sim T^{1+\nu}$, where nonlinearities appear in
the current-voltage characteristics decreases with temperature roughly
as $T^2$ in contrast with the $T^3$ behavior expected for disordered
models \cite{fisher2}. This could provide a signature of the behavior
at irrational frustration in experimental studies of ordered arrays of
Josephson junctions.

We considered an ordered array of Josephson junctions defined
on a square lattice.  The I-V characteristics of the array was computed
using an overdamped Langevin molecular-dynamics as described by Falo et
al \cite{falo}. One allows for a capacitance to the ground $C_o$ in
addition to a shunt resistance $R_0$ between superconducting grains. In
the overdamped limit, this particular dynamics reduces to the standard
resistance shunt junction (RSJ) model \cite{shenoy} commonly used in
dynamical simulations \cite{stroud}. The Langevin equations can be
written as
\begin{equation}
C_o \frac{d^2 \theta_i}{d t^2} + 
\frac{1}{R_o}\sum_{j} \frac{d (\theta_i -\theta_j)}{d t} =
-I_c\sum_j \sin(\theta_i - \theta_j - A_{ij}) +I_i^{ ext}+ \sum_{j} \eta_{ij}
\end{equation}
where $\eta_{ij}$ is a Gaussian white noise satisfying
\begin{equation}
<\eta_{ij}(t) > = 0, \quad   
<\eta_{ij}(t)\eta_{kl}(t')>=\frac{2 k_B T}{R_o}\delta_{ij,kl}\delta(t-t')
\end{equation}
and $i$, $j$ are nearest-neighbor pairs.  $I^{ ext}$ is the external
current  and $I_c$  the Josephson-junction critical current. The bond
variables  $A_{ij}$ are constrained to $\sum_{ij} = 2 \pi f$ where
$\sum_{ij}$ represent a direct sum  around  a plaquette.  We use units
where $\hbar/2e =1$,$R_o=1$, $I_c=1$ and set the parameter $I_c R_o^2
C_o  = 0.5$ in the simulations, corresponding to the overdamped
regime.  To determine the nonequilibrium voltage across the system, an
external current $I$ is injected uniformly with density $J = I/L$ along
one edge of a square array of size $L \times L$ and extracted at the
opposite one. Periodic boundary conditions are used in the transverse
direction to the current. The average voltage  drop $V$ across the
system is given by
\begin{equation}
V= \frac{1}{L} \frac{\hbar}{2e} \sum_{j=1}^{L}<\frac{d \theta_{1,j}}{dt} - 
\frac{d\theta_{L,j}}{dt} >
\end{equation}
and the average electric field  by $E = V/L$. The dynamical equations
were integrated numerically using typically a time step $\delta t =
0.03 \tau$ ($\tau = \hbar/(2 e R_o I_c)$) and averages computed with
$2 \times 10^5$ time steps for each calculation. Lattices of sizes
$L=21$ and $L=34$ were used in the simulations without significant size
dependence.

In Fig. 1, we show the nonlinear I-V characteristics as function of
temperature. At small current densities $J$, the nonlinear resistivity
$E/J$ shows a linear contribution where $E/J$ is a constant.  The
linear resistivity decreases rapidly with decreasing temperature.  For
increasing $J$, the resistivity cross over to a nonlinear behavior at a
critical current which also decreases with decreasing temperature.  For
$T < 0.25$, the linear behavior presumably occurs at current densities
much smaller than the lowest value used in the simulations, $J=0.02$.
However, in order to confirm this behavior a much better statistics and
long equilibration times would be required to obtain reliable data.
This is consistent with a kind of frozen state for temperatures below
$T \sim 0.25$ as indicated in Monte Carlo  simulations \cite{halsey}
where a nonzero Edwards-Anderson order parameter $q(t) = <\vec S_i>^2$,
where $\vec S =(\cos(\theta),\sin(\theta))$, averaged over long times
$t$, was observed below this temperature. This implies a diverging
relaxation time $\tau \sim  \int_0^\infty q(t) d t$ which prevents one
to obtain fully equilibrated data in this low temperature regime.

In Fig. 2 we show the behavior of the linear resistance as a function
of temperature, estimated from $E/J$ at the smallest current. It
appears consistent with an activated Arrhenius behavior with an
estimated energy barrier $E_b \sim 0.45$. We note that this energy
barrier is higher than the single vortex barrier in a periodic array
\cite{lobb}, $E_b \sim 0.19$.  This suggests that many vortices are
involved in the activation process and the vortex dynamics could be the
result of  collective motion of correlated vortices within a finite
length scale.

The behavior of the nonlinear I-V characteristics described above can
be understood within a scaling theory which assumes a second-order
phase transition at zero temperature \cite{fisher1}. This theory has
been applied to the study of models of vortex glasses in disordered
two-dimensional superconductors \cite{fisher2,koch} but it can be equally
applied to our case once a zero temperature transition is accepted.
Basically,  the behavior of the I-V curves is strongly affected by the
increasing correlation length $\xi$ which is assumed to diverge at zero
temperature as $\xi \sim T^{-\nu}$ with a thermal correlation exponent
$\nu$. At small current densities $J$, the behavior is dominated by
vortex dynamics at length scales larger than $\xi$ where the system
behaves as a vortex liquid with a linear resistance $R_L$.  At higher
currents, one probes length scales smaller than $\xi$ where the vortex
dynamics is expected to be nonlinear.  Near the transition, the
nonlinear  I-V characteristics can be cast into a scaling form
\cite{fisher1,fisher2}
\begin{equation}
\frac{E}{JR_L} = g (\frac{J}{T^{\nu +1}})
\end{equation}
where $g$ is a scaling function which has the property that $g(x)
\rightarrow 1$ when $x \rightarrow 0$.  Since at nonzero temperatures
the correlation length is finite, vortex motion will proceed by thermal
activation leading to a linear resistance in the form
\begin{equation}
R_L \propto \exp(-E_b/T)
\end{equation}
where $E_b$ is a typical barrier height for vortex motion within a
correlation length $\xi$. In general, the barrier energy should also
scale  with $\xi$ as  $E_b \sim \xi^\psi$ with a new exponent $\psi \ge
0$ but the present data do not allow a clear identification of this
behavior. The Arrhenius plot in Fig. 2 is consistent with the simplest
scenario $\psi =0$ with $E_b \sim 0.45$. The characteristic time scale
$\tau$ should also have an exponential behavior $\tau \sim \exp (E_b/
T)$ leading to very slow relaxation at low temperatures consistent with
the results above and Monte Carlo simulations \cite{halsey}.
 The estimated $E_b$ being larger than the
single vortex pinning potential \cite{lobb} $E_p=0.19$  is also
consistent with a relaxation process involving collective rearrangement
of pinned vortices.
From the scaling behavior above, one also sees
that the characteristic current density at which nonlinear behavior is
expected to set in varies as $J_{nl} \sim T^{1+\nu}$. In a finite
system, there will also be a dependence on the system size but the
system sizes we have used, $L=21$ and $L=34$, do not show significant 
corrections.  

In Fig.  3, we show a scaling plot of the I-V characteristics data of
Fig. 1. From the above scaling analysis all data should collapse on to
the same curve if $\nu$ is chosen correctly, assuming size corrections
are small. Choosing different values of $\nu$, we find that a
reasonable scaling behavior is obtained for $\nu \sim 0.9 \pm 0.2$.
This value of $\nu$ should be compared with the corresponding value for
the gauge glass model \cite{fisher2} of disordered superconductors in a
magnetic field, $\nu \sim 2$.  As expected, the different nature of the
two vortex glass transitions is reflected in the values of this
critical exponent. The sharp difference between the critical exponents
also leads to a completely different behavior for the crossover current
density which is expected to behave as $J_{nl} \sim T^2$ in contrast
with the $\sim T^3$ behavior for models of vortex glasses in disordered
superconductors. In the experimental systems additional complications
may arise since a higher order rational value can always be found
within the confidence interval of a measured value of $f$.

In conclusion, the current-voltage characteristics of an ordered
Josephson junction array at an irrational flux quantum per plaquette
was studied using an scaling analysis of a zero temperature transition.
The  correlation critical exponent is found to be significantly
different from the corresponding value for  vortex-glass models in
disordered two-dimensional superconductors.  This leads to a current
scale where nonlinearities sets in decreasing with temperature roughly as
$T^2$ in contrast with the $T^3$ behavior expected for disordered
models. It is suggested that this behavior could provide an
experimental signature for an array at irrational frustration.


\newpage

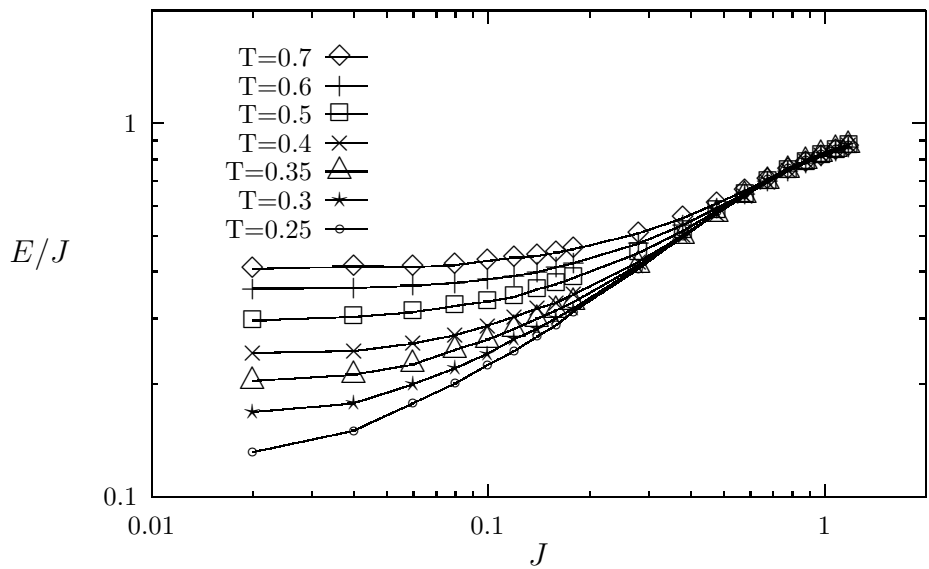
\begin{figure}
\caption{Nonlinear current-voltage characteristics at different temperatures  
for a system size $L=34$. }
\end{figure}

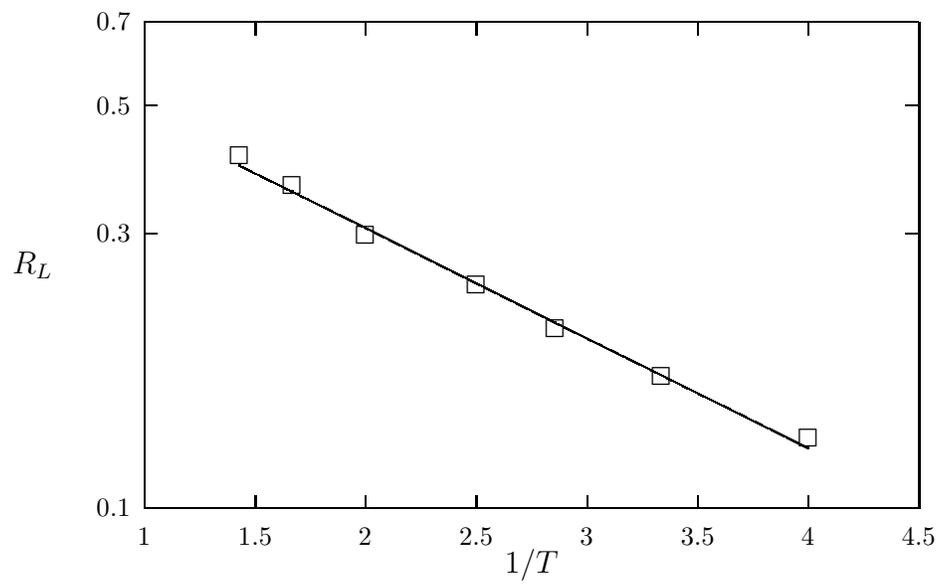
\begin{figure}
\caption{Temperature dependence of the linear resistance as estimated
from $E/J$ at $J=0.02$.}
\end{figure}

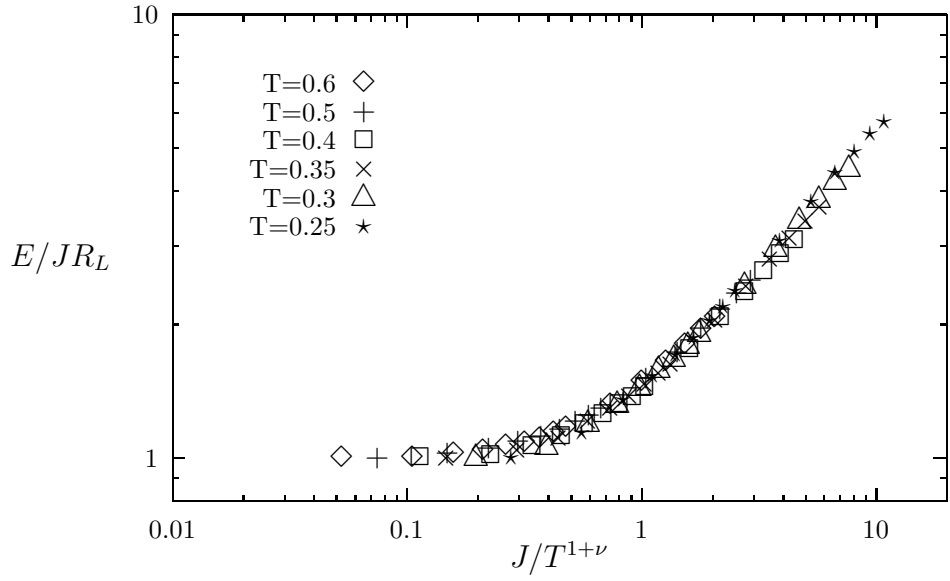
\begin{figure}
\caption{Scaling plot of the nonlinear current-voltage characteristics of Fig. 1
for $\nu = 0.9$.}
\end{figure}

\newpage
\begin{figure}
\begin{center}

\setlength{\unitlength}{0.240900pt}
\ifx\plotpoint\undefined\newsavebox{\plotpoint}\fi
\sbox{\plotpoint}{\rule[-0.200pt]{0.400pt}{0.400pt}}%
\begin{picture}(1500,900)(0,0)
\font\gnuplot=cmr10 at 10pt
\gnuplot
\sbox{\plotpoint}{\rule[-0.200pt]{0.400pt}{0.400pt}}%
\put(220.0,113.0){\rule[-0.200pt]{4.818pt}{0.400pt}}
\put(198,113){\makebox(0,0)[r]{0.1}}
\put(1416.0,113.0){\rule[-0.200pt]{4.818pt}{0.400pt}}
\put(220.0,290.0){\rule[-0.200pt]{2.409pt}{0.400pt}}
\put(1426.0,290.0){\rule[-0.200pt]{2.409pt}{0.400pt}}
\put(220.0,393.0){\rule[-0.200pt]{2.409pt}{0.400pt}}
\put(1426.0,393.0){\rule[-0.200pt]{2.409pt}{0.400pt}}
\put(220.0,467.0){\rule[-0.200pt]{2.409pt}{0.400pt}}
\put(1426.0,467.0){\rule[-0.200pt]{2.409pt}{0.400pt}}
\put(220.0,523.0){\rule[-0.200pt]{2.409pt}{0.400pt}}
\put(1426.0,523.0){\rule[-0.200pt]{2.409pt}{0.400pt}}
\put(220.0,570.0){\rule[-0.200pt]{2.409pt}{0.400pt}}
\put(1426.0,570.0){\rule[-0.200pt]{2.409pt}{0.400pt}}
\put(220.0,609.0){\rule[-0.200pt]{2.409pt}{0.400pt}}
\put(1426.0,609.0){\rule[-0.200pt]{2.409pt}{0.400pt}}
\put(220.0,643.0){\rule[-0.200pt]{2.409pt}{0.400pt}}
\put(1426.0,643.0){\rule[-0.200pt]{2.409pt}{0.400pt}}
\put(220.0,673.0){\rule[-0.200pt]{2.409pt}{0.400pt}}
\put(1426.0,673.0){\rule[-0.200pt]{2.409pt}{0.400pt}}
\put(220.0,700.0){\rule[-0.200pt]{4.818pt}{0.400pt}}
\put(198,700){\makebox(0,0)[r]{1}}
\put(1416.0,700.0){\rule[-0.200pt]{4.818pt}{0.400pt}}
\put(220.0,877.0){\rule[-0.200pt]{2.409pt}{0.400pt}}
\put(1426.0,877.0){\rule[-0.200pt]{2.409pt}{0.400pt}}
\put(220.0,113.0){\rule[-0.200pt]{0.400pt}{4.818pt}}
\put(220,68){\makebox(0,0){0.01}}
\put(220.0,857.0){\rule[-0.200pt]{0.400pt}{4.818pt}}
\put(379.0,113.0){\rule[-0.200pt]{0.400pt}{2.409pt}}
\put(379.0,867.0){\rule[-0.200pt]{0.400pt}{2.409pt}}
\put(472.0,113.0){\rule[-0.200pt]{0.400pt}{2.409pt}}
\put(472.0,867.0){\rule[-0.200pt]{0.400pt}{2.409pt}}
\put(538.0,113.0){\rule[-0.200pt]{0.400pt}{2.409pt}}
\put(538.0,867.0){\rule[-0.200pt]{0.400pt}{2.409pt}}
\put(589.0,113.0){\rule[-0.200pt]{0.400pt}{2.409pt}}
\put(589.0,867.0){\rule[-0.200pt]{0.400pt}{2.409pt}}
\put(631.0,113.0){\rule[-0.200pt]{0.400pt}{2.409pt}}
\put(631.0,867.0){\rule[-0.200pt]{0.400pt}{2.409pt}}
\put(667.0,113.0){\rule[-0.200pt]{0.400pt}{2.409pt}}
\put(667.0,867.0){\rule[-0.200pt]{0.400pt}{2.409pt}}
\put(697.0,113.0){\rule[-0.200pt]{0.400pt}{2.409pt}}
\put(697.0,867.0){\rule[-0.200pt]{0.400pt}{2.409pt}}
\put(724.0,113.0){\rule[-0.200pt]{0.400pt}{2.409pt}}
\put(724.0,867.0){\rule[-0.200pt]{0.400pt}{2.409pt}}
\put(748.0,113.0){\rule[-0.200pt]{0.400pt}{4.818pt}}
\put(748,68){\makebox(0,0){0.1}}
\put(748.0,857.0){\rule[-0.200pt]{0.400pt}{4.818pt}}
\put(908.0,113.0){\rule[-0.200pt]{0.400pt}{2.409pt}}
\put(908.0,867.0){\rule[-0.200pt]{0.400pt}{2.409pt}}
\put(1001.0,113.0){\rule[-0.200pt]{0.400pt}{2.409pt}}
\put(1001.0,867.0){\rule[-0.200pt]{0.400pt}{2.409pt}}
\put(1067.0,113.0){\rule[-0.200pt]{0.400pt}{2.409pt}}
\put(1067.0,867.0){\rule[-0.200pt]{0.400pt}{2.409pt}}
\put(1118.0,113.0){\rule[-0.200pt]{0.400pt}{2.409pt}}
\put(1118.0,867.0){\rule[-0.200pt]{0.400pt}{2.409pt}}
\put(1160.0,113.0){\rule[-0.200pt]{0.400pt}{2.409pt}}
\put(1160.0,867.0){\rule[-0.200pt]{0.400pt}{2.409pt}}
\put(1195.0,113.0){\rule[-0.200pt]{0.400pt}{2.409pt}}
\put(1195.0,867.0){\rule[-0.200pt]{0.400pt}{2.409pt}}
\put(1226.0,113.0){\rule[-0.200pt]{0.400pt}{2.409pt}}
\put(1226.0,867.0){\rule[-0.200pt]{0.400pt}{2.409pt}}
\put(1253.0,113.0){\rule[-0.200pt]{0.400pt}{2.409pt}}
\put(1253.0,867.0){\rule[-0.200pt]{0.400pt}{2.409pt}}
\put(1277.0,113.0){\rule[-0.200pt]{0.400pt}{4.818pt}}
\put(1277,68){\makebox(0,0){1}}
\put(1277.0,857.0){\rule[-0.200pt]{0.400pt}{4.818pt}}
\put(1436.0,113.0){\rule[-0.200pt]{0.400pt}{2.409pt}}
\put(1436.0,867.0){\rule[-0.200pt]{0.400pt}{2.409pt}}
\put(220.0,113.0){\rule[-0.200pt]{292.934pt}{0.400pt}}
\put(1436.0,113.0){\rule[-0.200pt]{0.400pt}{184.048pt}}
\put(220.0,877.0){\rule[-0.200pt]{292.934pt}{0.400pt}}
\put(45,495){\makebox(0,0){$E/J$}}
\put(828,23){\makebox(0,0){$J$}}
\put(220.0,113.0){\rule[-0.200pt]{0.400pt}{184.048pt}}
\put(472,804){\makebox(0,0)[r]{T=0.7}}
\put(494.0,804.0){\rule[-0.200pt]{15.899pt}{0.400pt}}
\put(379,471){\usebox{\plotpoint}}
\multiput(379.00,471.60)(23.146,0.468){5}{\rule{16.000pt}{0.113pt}}
\multiput(379.00,470.17)(125.791,4.000){2}{\rule{8.000pt}{0.400pt}}
\put(538,473.67){\rule{22.404pt}{0.400pt}}
\multiput(538.00,474.17)(46.500,-1.000){2}{\rule{11.202pt}{0.400pt}}
\multiput(631.00,474.61)(14.528,0.447){3}{\rule{8.900pt}{0.108pt}}
\multiput(631.00,473.17)(47.528,3.000){2}{\rule{4.450pt}{0.400pt}}
\multiput(697.00,477.59)(3.849,0.485){11}{\rule{3.014pt}{0.117pt}}
\multiput(697.00,476.17)(44.744,7.000){2}{\rule{1.507pt}{0.400pt}}
\multiput(748.00,484.59)(4.606,0.477){7}{\rule{3.460pt}{0.115pt}}
\multiput(748.00,483.17)(34.819,5.000){2}{\rule{1.730pt}{0.400pt}}
\put(790,489.17){\rule{7.300pt}{0.400pt}}
\multiput(790.00,488.17)(20.848,2.000){2}{\rule{3.650pt}{0.400pt}}
\multiput(826.00,491.59)(2.660,0.482){9}{\rule{2.100pt}{0.116pt}}
\multiput(826.00,490.17)(25.641,6.000){2}{\rule{1.050pt}{0.400pt}}
\multiput(856.00,497.59)(2.936,0.477){7}{\rule{2.260pt}{0.115pt}}
\multiput(856.00,496.17)(22.309,5.000){2}{\rule{1.130pt}{0.400pt}}
\multiput(883.00,502.58)(2.061,0.497){47}{\rule{1.732pt}{0.120pt}}
\multiput(883.00,501.17)(98.405,25.000){2}{\rule{0.866pt}{0.400pt}}
\multiput(985.00,527.58)(1.471,0.496){45}{\rule{1.267pt}{0.120pt}}
\multiput(985.00,526.17)(67.371,24.000){2}{\rule{0.633pt}{0.400pt}}
\multiput(1055.00,551.58)(1.161,0.496){43}{\rule{1.022pt}{0.120pt}}
\multiput(1055.00,550.17)(50.879,23.000){2}{\rule{0.511pt}{0.400pt}}
\multiput(1108.00,574.58)(1.109,0.496){37}{\rule{0.980pt}{0.119pt}}
\multiput(1108.00,573.17)(41.966,20.000){2}{\rule{0.490pt}{0.400pt}}
\multiput(1152.00,594.58)(1.069,0.495){31}{\rule{0.947pt}{0.119pt}}
\multiput(1152.00,593.17)(34.034,17.000){2}{\rule{0.474pt}{0.400pt}}
\multiput(1188.00,611.58)(1.079,0.494){27}{\rule{0.953pt}{0.119pt}}
\multiput(1188.00,610.17)(30.021,15.000){2}{\rule{0.477pt}{0.400pt}}
\multiput(1220.00,626.58)(1.186,0.492){21}{\rule{1.033pt}{0.119pt}}
\multiput(1220.00,625.17)(25.855,12.000){2}{\rule{0.517pt}{0.400pt}}
\multiput(1248.00,638.58)(1.225,0.491){17}{\rule{1.060pt}{0.118pt}}
\multiput(1248.00,637.17)(21.800,10.000){2}{\rule{0.530pt}{0.400pt}}
\multiput(1272.00,648.59)(1.484,0.488){13}{\rule{1.250pt}{0.117pt}}
\multiput(1272.00,647.17)(20.406,8.000){2}{\rule{0.625pt}{0.400pt}}
\multiput(1295.00,656.59)(1.484,0.485){11}{\rule{1.243pt}{0.117pt}}
\multiput(1295.00,655.17)(17.420,7.000){2}{\rule{0.621pt}{0.400pt}}
\put(516,804){\raisebox{-.8pt}{\makebox(0,0){$\Diamond$}}}
\put(379,471){\raisebox{-.8pt}{\makebox(0,0){$\Diamond$}}}
\put(538,475){\raisebox{-.8pt}{\makebox(0,0){$\Diamond$}}}
\put(631,474){\raisebox{-.8pt}{\makebox(0,0){$\Diamond$}}}
\put(697,477){\raisebox{-.8pt}{\makebox(0,0){$\Diamond$}}}
\put(748,484){\raisebox{-.8pt}{\makebox(0,0){$\Diamond$}}}
\put(790,489){\raisebox{-.8pt}{\makebox(0,0){$\Diamond$}}}
\put(826,491){\raisebox{-.8pt}{\makebox(0,0){$\Diamond$}}}
\put(856,497){\raisebox{-.8pt}{\makebox(0,0){$\Diamond$}}}
\put(883,502){\raisebox{-.8pt}{\makebox(0,0){$\Diamond$}}}
\put(985,527){\raisebox{-.8pt}{\makebox(0,0){$\Diamond$}}}
\put(1055,551){\raisebox{-.8pt}{\makebox(0,0){$\Diamond$}}}
\put(1108,574){\raisebox{-.8pt}{\makebox(0,0){$\Diamond$}}}
\put(1152,594){\raisebox{-.8pt}{\makebox(0,0){$\Diamond$}}}
\put(1188,611){\raisebox{-.8pt}{\makebox(0,0){$\Diamond$}}}
\put(1220,626){\raisebox{-.8pt}{\makebox(0,0){$\Diamond$}}}
\put(1248,638){\raisebox{-.8pt}{\makebox(0,0){$\Diamond$}}}
\put(1272,648){\raisebox{-.8pt}{\makebox(0,0){$\Diamond$}}}
\put(1295,656){\raisebox{-.8pt}{\makebox(0,0){$\Diamond$}}}
\put(1315,663){\raisebox{-.8pt}{\makebox(0,0){$\Diamond$}}}
\put(472,759){\makebox(0,0)[r]{T=0.6}}
\put(494.0,759.0){\rule[-0.200pt]{15.899pt}{0.400pt}}
\put(379,440){\usebox{\plotpoint}}
\put(379,439.67){\rule{38.303pt}{0.400pt}}
\multiput(379.00,439.17)(79.500,1.000){2}{\rule{19.152pt}{0.400pt}}
\multiput(538.00,441.60)(13.495,0.468){5}{\rule{9.400pt}{0.113pt}}
\multiput(538.00,440.17)(73.490,4.000){2}{\rule{4.700pt}{0.400pt}}
\multiput(631.00,445.60)(9.547,0.468){5}{\rule{6.700pt}{0.113pt}}
\multiput(631.00,444.17)(52.094,4.000){2}{\rule{3.350pt}{0.400pt}}
\multiput(697.00,449.59)(4.559,0.482){9}{\rule{3.500pt}{0.116pt}}
\multiput(697.00,448.17)(43.736,6.000){2}{\rule{1.750pt}{0.400pt}}
\multiput(748.00,455.59)(4.606,0.477){7}{\rule{3.460pt}{0.115pt}}
\multiput(748.00,454.17)(34.819,5.000){2}{\rule{1.730pt}{0.400pt}}
\multiput(790.00,460.59)(3.203,0.482){9}{\rule{2.500pt}{0.116pt}}
\multiput(790.00,459.17)(30.811,6.000){2}{\rule{1.250pt}{0.400pt}}
\multiput(826.00,466.59)(1.947,0.488){13}{\rule{1.600pt}{0.117pt}}
\multiput(826.00,465.17)(26.679,8.000){2}{\rule{0.800pt}{0.400pt}}
\multiput(856.00,474.59)(2.389,0.482){9}{\rule{1.900pt}{0.116pt}}
\multiput(856.00,473.17)(23.056,6.000){2}{\rule{0.950pt}{0.400pt}}
\multiput(883.00,480.58)(1.657,0.497){59}{\rule{1.416pt}{0.120pt}}
\multiput(883.00,479.17)(99.061,31.000){2}{\rule{0.708pt}{0.400pt}}
\multiput(985.00,511.58)(1.214,0.497){55}{\rule{1.066pt}{0.120pt}}
\multiput(985.00,510.17)(67.788,29.000){2}{\rule{0.533pt}{0.400pt}}
\multiput(1055.00,540.58)(0.987,0.497){51}{\rule{0.885pt}{0.120pt}}
\multiput(1055.00,539.17)(51.163,27.000){2}{\rule{0.443pt}{0.400pt}}
\multiput(1108.00,567.58)(1.007,0.496){41}{\rule{0.900pt}{0.120pt}}
\multiput(1108.00,566.17)(42.132,22.000){2}{\rule{0.450pt}{0.400pt}}
\multiput(1152.00,589.58)(0.905,0.496){37}{\rule{0.820pt}{0.119pt}}
\multiput(1152.00,588.17)(34.298,20.000){2}{\rule{0.410pt}{0.400pt}}
\multiput(1188.00,609.58)(1.009,0.494){29}{\rule{0.900pt}{0.119pt}}
\multiput(1188.00,608.17)(30.132,16.000){2}{\rule{0.450pt}{0.400pt}}
\multiput(1220.00,625.58)(1.091,0.493){23}{\rule{0.962pt}{0.119pt}}
\multiput(1220.00,624.17)(26.004,13.000){2}{\rule{0.481pt}{0.400pt}}
\multiput(1248.00,638.58)(1.225,0.491){17}{\rule{1.060pt}{0.118pt}}
\multiput(1248.00,637.17)(21.800,10.000){2}{\rule{0.530pt}{0.400pt}}
\multiput(1272.00,648.59)(1.484,0.488){13}{\rule{1.250pt}{0.117pt}}
\multiput(1272.00,647.17)(20.406,8.000){2}{\rule{0.625pt}{0.400pt}}
\multiput(1295.00,656.59)(1.484,0.485){11}{\rule{1.243pt}{0.117pt}}
\multiput(1295.00,655.17)(17.420,7.000){2}{\rule{0.621pt}{0.400pt}}
\put(516,759){\makebox(0,0){$+$}}
\put(379,440){\makebox(0,0){$+$}}
\put(538,441){\makebox(0,0){$+$}}
\put(631,445){\makebox(0,0){$+$}}
\put(697,449){\makebox(0,0){$+$}}
\put(748,455){\makebox(0,0){$+$}}
\put(790,460){\makebox(0,0){$+$}}
\put(826,466){\makebox(0,0){$+$}}
\put(856,474){\makebox(0,0){$+$}}
\put(883,480){\makebox(0,0){$+$}}
\put(985,511){\makebox(0,0){$+$}}
\put(1055,540){\makebox(0,0){$+$}}
\put(1108,567){\makebox(0,0){$+$}}
\put(1152,589){\makebox(0,0){$+$}}
\put(1188,609){\makebox(0,0){$+$}}
\put(1220,625){\makebox(0,0){$+$}}
\put(1248,638){\makebox(0,0){$+$}}
\put(1272,648){\makebox(0,0){$+$}}
\put(1295,656){\makebox(0,0){$+$}}
\put(1315,663){\makebox(0,0){$+$}}
\put(472,714){\makebox(0,0)[r]{T=0.5}}
\put(494.0,714.0){\rule[-0.200pt]{15.899pt}{0.400pt}}
\put(379,390){\usebox{\plotpoint}}
\multiput(379.00,390.59)(14.323,0.482){9}{\rule{10.700pt}{0.116pt}}
\multiput(379.00,389.17)(136.792,6.000){2}{\rule{5.350pt}{0.400pt}}
\multiput(538.00,396.59)(7.052,0.485){11}{\rule{5.414pt}{0.117pt}}
\multiput(538.00,395.17)(81.762,7.000){2}{\rule{2.707pt}{0.400pt}}
\multiput(631.00,403.58)(3.414,0.491){17}{\rule{2.740pt}{0.118pt}}
\multiput(631.00,402.17)(60.313,10.000){2}{\rule{1.370pt}{0.400pt}}
\multiput(697.00,413.59)(3.849,0.485){11}{\rule{3.014pt}{0.117pt}}
\multiput(697.00,412.17)(44.744,7.000){2}{\rule{1.507pt}{0.400pt}}
\multiput(748.00,420.59)(3.162,0.485){11}{\rule{2.500pt}{0.117pt}}
\multiput(748.00,419.17)(36.811,7.000){2}{\rule{1.250pt}{0.400pt}}
\multiput(790.00,427.58)(1.530,0.492){21}{\rule{1.300pt}{0.119pt}}
\multiput(790.00,426.17)(33.302,12.000){2}{\rule{0.650pt}{0.400pt}}
\multiput(826.00,439.59)(1.947,0.488){13}{\rule{1.600pt}{0.117pt}}
\multiput(826.00,438.17)(26.679,8.000){2}{\rule{0.800pt}{0.400pt}}
\multiput(856.00,447.58)(1.381,0.491){17}{\rule{1.180pt}{0.118pt}}
\multiput(856.00,446.17)(24.551,10.000){2}{\rule{0.590pt}{0.400pt}}
\multiput(883.00,457.58)(1.314,0.498){75}{\rule{1.146pt}{0.120pt}}
\multiput(883.00,456.17)(99.621,39.000){2}{\rule{0.573pt}{0.400pt}}
\multiput(985.00,496.58)(1.004,0.498){67}{\rule{0.900pt}{0.120pt}}
\multiput(985.00,495.17)(68.132,35.000){2}{\rule{0.450pt}{0.400pt}}
\multiput(1055.00,531.58)(0.858,0.497){59}{\rule{0.784pt}{0.120pt}}
\multiput(1055.00,530.17)(51.373,31.000){2}{\rule{0.392pt}{0.400pt}}
\multiput(1108.00,562.58)(0.884,0.497){47}{\rule{0.804pt}{0.120pt}}
\multiput(1108.00,561.17)(42.331,25.000){2}{\rule{0.402pt}{0.400pt}}
\multiput(1152.00,587.58)(0.862,0.496){39}{\rule{0.786pt}{0.119pt}}
\multiput(1152.00,586.17)(34.369,21.000){2}{\rule{0.393pt}{0.400pt}}
\multiput(1188.00,608.58)(0.949,0.495){31}{\rule{0.853pt}{0.119pt}}
\multiput(1188.00,607.17)(30.230,17.000){2}{\rule{0.426pt}{0.400pt}}
\multiput(1220.00,625.58)(1.091,0.493){23}{\rule{0.962pt}{0.119pt}}
\multiput(1220.00,624.17)(26.004,13.000){2}{\rule{0.481pt}{0.400pt}}
\multiput(1248.00,638.58)(1.109,0.492){19}{\rule{0.973pt}{0.118pt}}
\multiput(1248.00,637.17)(21.981,11.000){2}{\rule{0.486pt}{0.400pt}}
\multiput(1272.00,649.59)(1.484,0.488){13}{\rule{1.250pt}{0.117pt}}
\multiput(1272.00,648.17)(20.406,8.000){2}{\rule{0.625pt}{0.400pt}}
\multiput(1295.00,657.59)(1.484,0.485){11}{\rule{1.243pt}{0.117pt}}
\multiput(1295.00,656.17)(17.420,7.000){2}{\rule{0.621pt}{0.400pt}}
\put(516,714){\raisebox{-.8pt}{\makebox(0,0){$\Box$}}}
\put(379,390){\raisebox{-.8pt}{\makebox(0,0){$\Box$}}}
\put(538,396){\raisebox{-.8pt}{\makebox(0,0){$\Box$}}}
\put(631,403){\raisebox{-.8pt}{\makebox(0,0){$\Box$}}}
\put(697,413){\raisebox{-.8pt}{\makebox(0,0){$\Box$}}}
\put(748,420){\raisebox{-.8pt}{\makebox(0,0){$\Box$}}}
\put(790,427){\raisebox{-.8pt}{\makebox(0,0){$\Box$}}}
\put(826,439){\raisebox{-.8pt}{\makebox(0,0){$\Box$}}}
\put(856,447){\raisebox{-.8pt}{\makebox(0,0){$\Box$}}}
\put(883,457){\raisebox{-.8pt}{\makebox(0,0){$\Box$}}}
\put(985,496){\raisebox{-.8pt}{\makebox(0,0){$\Box$}}}
\put(1055,531){\raisebox{-.8pt}{\makebox(0,0){$\Box$}}}
\put(1108,562){\raisebox{-.8pt}{\makebox(0,0){$\Box$}}}
\put(1152,587){\raisebox{-.8pt}{\makebox(0,0){$\Box$}}}
\put(1188,608){\raisebox{-.8pt}{\makebox(0,0){$\Box$}}}
\put(1220,625){\raisebox{-.8pt}{\makebox(0,0){$\Box$}}}
\put(1248,638){\raisebox{-.8pt}{\makebox(0,0){$\Box$}}}
\put(1272,649){\raisebox{-.8pt}{\makebox(0,0){$\Box$}}}
\put(1295,657){\raisebox{-.8pt}{\makebox(0,0){$\Box$}}}
\put(1315,664){\raisebox{-.8pt}{\makebox(0,0){$\Box$}}}
\put(472,669){\makebox(0,0)[r]{T=0.4}}
\put(494.0,669.0){\rule[-0.200pt]{15.899pt}{0.400pt}}
\put(379,339){\usebox{\plotpoint}}
\multiput(379.00,339.61)(35.291,0.447){3}{\rule{21.300pt}{0.108pt}}
\multiput(379.00,338.17)(114.791,3.000){2}{\rule{10.650pt}{0.400pt}}
\multiput(538.00,342.58)(3.986,0.492){21}{\rule{3.200pt}{0.119pt}}
\multiput(538.00,341.17)(86.358,12.000){2}{\rule{1.600pt}{0.400pt}}
\multiput(631.00,354.58)(2.598,0.493){23}{\rule{2.131pt}{0.119pt}}
\multiput(631.00,353.17)(61.577,13.000){2}{\rule{1.065pt}{0.400pt}}
\multiput(697.00,367.58)(1.728,0.494){27}{\rule{1.460pt}{0.119pt}}
\multiput(697.00,366.17)(47.970,15.000){2}{\rule{0.730pt}{0.400pt}}
\multiput(748.00,382.58)(1.525,0.494){25}{\rule{1.300pt}{0.119pt}}
\multiput(748.00,381.17)(39.302,14.000){2}{\rule{0.650pt}{0.400pt}}
\multiput(790.00,396.58)(1.408,0.493){23}{\rule{1.208pt}{0.119pt}}
\multiput(790.00,395.17)(33.493,13.000){2}{\rule{0.604pt}{0.400pt}}
\multiput(826.00,409.59)(1.718,0.489){15}{\rule{1.433pt}{0.118pt}}
\multiput(826.00,408.17)(27.025,9.000){2}{\rule{0.717pt}{0.400pt}}
\multiput(856.00,418.58)(1.052,0.493){23}{\rule{0.931pt}{0.119pt}}
\multiput(856.00,417.17)(25.068,13.000){2}{\rule{0.465pt}{0.400pt}}
\multiput(883.00,431.58)(1.003,0.498){99}{\rule{0.900pt}{0.120pt}}
\multiput(883.00,430.17)(100.132,51.000){2}{\rule{0.450pt}{0.400pt}}
\multiput(985.00,482.58)(0.835,0.498){81}{\rule{0.767pt}{0.120pt}}
\multiput(985.00,481.17)(68.409,42.000){2}{\rule{0.383pt}{0.400pt}}
\multiput(1055.00,524.58)(0.781,0.498){65}{\rule{0.724pt}{0.120pt}}
\multiput(1055.00,523.17)(51.498,34.000){2}{\rule{0.362pt}{0.400pt}}
\multiput(1108.00,558.58)(0.788,0.497){53}{\rule{0.729pt}{0.120pt}}
\multiput(1108.00,557.17)(42.488,28.000){2}{\rule{0.364pt}{0.400pt}}
\multiput(1152.00,586.58)(0.822,0.496){41}{\rule{0.755pt}{0.120pt}}
\multiput(1152.00,585.17)(34.434,22.000){2}{\rule{0.377pt}{0.400pt}}
\multiput(1188.00,608.58)(0.895,0.495){33}{\rule{0.811pt}{0.119pt}}
\multiput(1188.00,607.17)(30.316,18.000){2}{\rule{0.406pt}{0.400pt}}
\multiput(1220.00,626.58)(1.011,0.494){25}{\rule{0.900pt}{0.119pt}}
\multiput(1220.00,625.17)(26.132,14.000){2}{\rule{0.450pt}{0.400pt}}
\multiput(1248.00,640.58)(1.225,0.491){17}{\rule{1.060pt}{0.118pt}}
\multiput(1248.00,639.17)(21.800,10.000){2}{\rule{0.530pt}{0.400pt}}
\multiput(1272.00,650.59)(1.484,0.488){13}{\rule{1.250pt}{0.117pt}}
\multiput(1272.00,649.17)(20.406,8.000){2}{\rule{0.625pt}{0.400pt}}
\multiput(1295.00,658.59)(1.484,0.485){11}{\rule{1.243pt}{0.117pt}}
\multiput(1295.00,657.17)(17.420,7.000){2}{\rule{0.621pt}{0.400pt}}
\put(516,669){\makebox(0,0){$\times$}}
\put(379,339){\makebox(0,0){$\times$}}
\put(538,342){\makebox(0,0){$\times$}}
\put(631,354){\makebox(0,0){$\times$}}
\put(697,367){\makebox(0,0){$\times$}}
\put(748,382){\makebox(0,0){$\times$}}
\put(790,396){\makebox(0,0){$\times$}}
\put(826,409){\makebox(0,0){$\times$}}
\put(856,418){\makebox(0,0){$\times$}}
\put(883,431){\makebox(0,0){$\times$}}
\put(985,482){\makebox(0,0){$\times$}}
\put(1055,524){\makebox(0,0){$\times$}}
\put(1108,558){\makebox(0,0){$\times$}}
\put(1152,586){\makebox(0,0){$\times$}}
\put(1188,608){\makebox(0,0){$\times$}}
\put(1220,626){\makebox(0,0){$\times$}}
\put(1248,640){\makebox(0,0){$\times$}}
\put(1272,650){\makebox(0,0){$\times$}}
\put(1295,658){\makebox(0,0){$\times$}}
\put(1315,665){\makebox(0,0){$\times$}}
\put(472,624){\makebox(0,0)[r]{T=0.35}}
\put(494.0,624.0){\rule[-0.200pt]{15.899pt}{0.400pt}}
\put(379,295){\usebox{\plotpoint}}
\multiput(379.00,295.58)(8.261,0.491){17}{\rule{6.460pt}{0.118pt}}
\multiput(379.00,294.17)(145.592,10.000){2}{\rule{3.230pt}{0.400pt}}
\multiput(538.00,305.58)(2.962,0.494){29}{\rule{2.425pt}{0.119pt}}
\multiput(538.00,304.17)(87.967,16.000){2}{\rule{1.213pt}{0.400pt}}
\multiput(631.00,321.58)(1.448,0.496){43}{\rule{1.248pt}{0.120pt}}
\multiput(631.00,320.17)(63.410,23.000){2}{\rule{0.624pt}{0.400pt}}
\multiput(697.00,344.58)(1.618,0.494){29}{\rule{1.375pt}{0.119pt}}
\multiput(697.00,343.17)(48.146,16.000){2}{\rule{0.688pt}{0.400pt}}
\multiput(748.00,360.58)(1.249,0.495){31}{\rule{1.088pt}{0.119pt}}
\multiput(748.00,359.17)(39.741,17.000){2}{\rule{0.544pt}{0.400pt}}
\multiput(790.00,377.58)(1.137,0.494){29}{\rule{1.000pt}{0.119pt}}
\multiput(790.00,376.17)(33.924,16.000){2}{\rule{0.500pt}{0.400pt}}
\multiput(826.00,393.58)(1.171,0.493){23}{\rule{1.023pt}{0.119pt}}
\multiput(826.00,392.17)(27.877,13.000){2}{\rule{0.512pt}{0.400pt}}
\multiput(856.00,406.58)(1.052,0.493){23}{\rule{0.931pt}{0.119pt}}
\multiput(856.00,405.17)(25.068,13.000){2}{\rule{0.465pt}{0.400pt}}
\multiput(883.00,419.58)(0.881,0.499){113}{\rule{0.803pt}{0.120pt}}
\multiput(883.00,418.17)(100.332,58.000){2}{\rule{0.402pt}{0.400pt}}
\multiput(985.00,477.58)(0.762,0.498){89}{\rule{0.709pt}{0.120pt}}
\multiput(985.00,476.17)(68.529,46.000){2}{\rule{0.354pt}{0.400pt}}
\multiput(1055.00,523.58)(0.759,0.498){67}{\rule{0.706pt}{0.120pt}}
\multiput(1055.00,522.17)(51.535,35.000){2}{\rule{0.353pt}{0.400pt}}
\multiput(1108.00,558.58)(0.760,0.497){55}{\rule{0.707pt}{0.120pt}}
\multiput(1108.00,557.17)(42.533,29.000){2}{\rule{0.353pt}{0.400pt}}
\multiput(1152.00,587.58)(0.822,0.496){41}{\rule{0.755pt}{0.120pt}}
\multiput(1152.00,586.17)(34.434,22.000){2}{\rule{0.377pt}{0.400pt}}
\multiput(1188.00,609.58)(0.895,0.495){33}{\rule{0.811pt}{0.119pt}}
\multiput(1188.00,608.17)(30.316,18.000){2}{\rule{0.406pt}{0.400pt}}
\multiput(1220.00,627.58)(1.011,0.494){25}{\rule{0.900pt}{0.119pt}}
\multiput(1220.00,626.17)(26.132,14.000){2}{\rule{0.450pt}{0.400pt}}
\multiput(1248.00,641.58)(1.225,0.491){17}{\rule{1.060pt}{0.118pt}}
\multiput(1248.00,640.17)(21.800,10.000){2}{\rule{0.530pt}{0.400pt}}
\multiput(1272.00,651.59)(1.484,0.488){13}{\rule{1.250pt}{0.117pt}}
\multiput(1272.00,650.17)(20.406,8.000){2}{\rule{0.625pt}{0.400pt}}
\multiput(1295.00,659.59)(1.756,0.482){9}{\rule{1.433pt}{0.116pt}}
\multiput(1295.00,658.17)(17.025,6.000){2}{\rule{0.717pt}{0.400pt}}
\put(516,624){\makebox(0,0){$\triangle$}}
\put(379,295){\makebox(0,0){$\triangle$}}
\put(538,305){\makebox(0,0){$\triangle$}}
\put(631,321){\makebox(0,0){$\triangle$}}
\put(697,344){\makebox(0,0){$\triangle$}}
\put(748,360){\makebox(0,0){$\triangle$}}
\put(790,377){\makebox(0,0){$\triangle$}}
\put(826,393){\makebox(0,0){$\triangle$}}
\put(856,406){\makebox(0,0){$\triangle$}}
\put(883,419){\makebox(0,0){$\triangle$}}
\put(985,477){\makebox(0,0){$\triangle$}}
\put(1055,523){\makebox(0,0){$\triangle$}}
\put(1108,558){\makebox(0,0){$\triangle$}}
\put(1152,587){\makebox(0,0){$\triangle$}}
\put(1188,609){\makebox(0,0){$\triangle$}}
\put(1220,627){\makebox(0,0){$\triangle$}}
\put(1248,641){\makebox(0,0){$\triangle$}}
\put(1272,651){\makebox(0,0){$\triangle$}}
\put(1295,659){\makebox(0,0){$\triangle$}}
\put(1315,665){\makebox(0,0){$\triangle$}}
\put(472,579){\makebox(0,0)[r]{T=0.3}}
\put(494.0,579.0){\rule[-0.200pt]{15.899pt}{0.400pt}}
\put(379,246){\usebox{\plotpoint}}
\multiput(379.00,246.58)(5.822,0.494){25}{\rule{4.643pt}{0.119pt}}
\multiput(379.00,245.17)(149.364,14.000){2}{\rule{2.321pt}{0.400pt}}
\multiput(538.00,260.58)(1.561,0.497){57}{\rule{1.340pt}{0.120pt}}
\multiput(538.00,259.17)(90.219,30.000){2}{\rule{0.670pt}{0.400pt}}
\multiput(631.00,290.58)(1.330,0.497){47}{\rule{1.156pt}{0.120pt}}
\multiput(631.00,289.17)(63.601,25.000){2}{\rule{0.578pt}{0.400pt}}
\multiput(697.00,315.58)(1.117,0.496){43}{\rule{0.987pt}{0.120pt}}
\multiput(697.00,314.17)(48.952,23.000){2}{\rule{0.493pt}{0.400pt}}
\multiput(748.00,338.58)(0.918,0.496){43}{\rule{0.830pt}{0.120pt}}
\multiput(748.00,337.17)(40.276,23.000){2}{\rule{0.415pt}{0.400pt}}
\multiput(790.00,361.58)(1.137,0.494){29}{\rule{1.000pt}{0.119pt}}
\multiput(790.00,360.17)(33.924,16.000){2}{\rule{0.500pt}{0.400pt}}
\multiput(826.00,377.58)(0.888,0.495){31}{\rule{0.806pt}{0.119pt}}
\multiput(826.00,376.17)(28.327,17.000){2}{\rule{0.403pt}{0.400pt}}
\multiput(856.00,394.58)(0.908,0.494){27}{\rule{0.820pt}{0.119pt}}
\multiput(856.00,393.17)(25.298,15.000){2}{\rule{0.410pt}{0.400pt}}
\multiput(883.00,409.58)(0.786,0.499){127}{\rule{0.728pt}{0.120pt}}
\multiput(883.00,408.17)(100.490,65.000){2}{\rule{0.364pt}{0.400pt}}
\multiput(985.00,474.58)(0.746,0.498){91}{\rule{0.696pt}{0.120pt}}
\multiput(985.00,473.17)(68.556,47.000){2}{\rule{0.348pt}{0.400pt}}
\multiput(1055.00,521.58)(0.698,0.498){73}{\rule{0.658pt}{0.120pt}}
\multiput(1055.00,520.17)(51.635,38.000){2}{\rule{0.329pt}{0.400pt}}
\multiput(1108.00,559.58)(0.760,0.497){55}{\rule{0.707pt}{0.120pt}}
\multiput(1108.00,558.17)(42.533,29.000){2}{\rule{0.353pt}{0.400pt}}
\multiput(1152.00,588.58)(0.785,0.496){43}{\rule{0.726pt}{0.120pt}}
\multiput(1152.00,587.17)(34.493,23.000){2}{\rule{0.363pt}{0.400pt}}
\multiput(1188.00,611.58)(0.949,0.495){31}{\rule{0.853pt}{0.119pt}}
\multiput(1188.00,610.17)(30.230,17.000){2}{\rule{0.426pt}{0.400pt}}
\multiput(1220.00,628.58)(1.011,0.494){25}{\rule{0.900pt}{0.119pt}}
\multiput(1220.00,627.17)(26.132,14.000){2}{\rule{0.450pt}{0.400pt}}
\multiput(1248.00,642.58)(1.225,0.491){17}{\rule{1.060pt}{0.118pt}}
\multiput(1248.00,641.17)(21.800,10.000){2}{\rule{0.530pt}{0.400pt}}
\multiput(1272.00,652.59)(1.484,0.488){13}{\rule{1.250pt}{0.117pt}}
\multiput(1272.00,651.17)(20.406,8.000){2}{\rule{0.625pt}{0.400pt}}
\multiput(1295.00,660.59)(1.756,0.482){9}{\rule{1.433pt}{0.116pt}}
\multiput(1295.00,659.17)(17.025,6.000){2}{\rule{0.717pt}{0.400pt}}
\put(516,579){\makebox(0,0){$\star$}}
\put(379,246){\makebox(0,0){$\star$}}
\put(538,260){\makebox(0,0){$\star$}}
\put(631,290){\makebox(0,0){$\star$}}
\put(697,315){\makebox(0,0){$\star$}}
\put(748,338){\makebox(0,0){$\star$}}
\put(790,361){\makebox(0,0){$\star$}}
\put(826,377){\makebox(0,0){$\star$}}
\put(856,394){\makebox(0,0){$\star$}}
\put(883,409){\makebox(0,0){$\star$}}
\put(985,474){\makebox(0,0){$\star$}}
\put(1055,521){\makebox(0,0){$\star$}}
\put(1108,559){\makebox(0,0){$\star$}}
\put(1152,588){\makebox(0,0){$\star$}}
\put(1188,611){\makebox(0,0){$\star$}}
\put(1220,628){\makebox(0,0){$\star$}}
\put(1248,642){\makebox(0,0){$\star$}}
\put(1272,652){\makebox(0,0){$\star$}}
\put(1295,660){\makebox(0,0){$\star$}}
\put(1315,666){\makebox(0,0){$\star$}}
\put(472,534){\makebox(0,0)[r]{T=0.25}}
\put(494.0,534.0){\rule[-0.200pt]{15.899pt}{0.400pt}}
\put(379,183){\usebox{\plotpoint}}
\multiput(379.00,183.58)(2.357,0.498){65}{\rule{1.971pt}{0.120pt}}
\multiput(379.00,182.17)(154.910,34.000){2}{\rule{0.985pt}{0.400pt}}
\multiput(538.00,217.58)(1.085,0.498){83}{\rule{0.965pt}{0.120pt}}
\multiput(538.00,216.17)(90.997,43.000){2}{\rule{0.483pt}{0.400pt}}
\multiput(631.00,260.58)(1.070,0.497){59}{\rule{0.952pt}{0.120pt}}
\multiput(631.00,259.17)(64.025,31.000){2}{\rule{0.476pt}{0.400pt}}
\multiput(697.00,291.58)(0.883,0.497){55}{\rule{0.803pt}{0.120pt}}
\multiput(697.00,290.17)(49.332,29.000){2}{\rule{0.402pt}{0.400pt}}
\multiput(748.00,320.58)(0.960,0.496){41}{\rule{0.864pt}{0.120pt}}
\multiput(748.00,319.17)(40.207,22.000){2}{\rule{0.432pt}{0.400pt}}
\multiput(790.00,342.58)(0.785,0.496){43}{\rule{0.726pt}{0.120pt}}
\multiput(790.00,341.17)(34.493,23.000){2}{\rule{0.363pt}{0.400pt}}
\multiput(826.00,365.58)(0.838,0.495){33}{\rule{0.767pt}{0.119pt}}
\multiput(826.00,364.17)(28.409,18.000){2}{\rule{0.383pt}{0.400pt}}
\multiput(856.00,383.58)(0.643,0.496){39}{\rule{0.614pt}{0.119pt}}
\multiput(856.00,382.17)(25.725,21.000){2}{\rule{0.307pt}{0.400pt}}
\multiput(883.00,404.58)(0.762,0.499){131}{\rule{0.709pt}{0.120pt}}
\multiput(883.00,403.17)(100.529,67.000){2}{\rule{0.354pt}{0.400pt}}
\multiput(985.00,471.58)(0.673,0.498){101}{\rule{0.638pt}{0.120pt}}
\multiput(985.00,470.17)(68.675,52.000){2}{\rule{0.319pt}{0.400pt}}
\multiput(1055.00,523.58)(0.717,0.498){71}{\rule{0.673pt}{0.120pt}}
\multiput(1055.00,522.17)(51.603,37.000){2}{\rule{0.336pt}{0.400pt}}
\multiput(1108.00,560.58)(0.760,0.497){55}{\rule{0.707pt}{0.120pt}}
\multiput(1108.00,559.17)(42.533,29.000){2}{\rule{0.353pt}{0.400pt}}
\multiput(1152.00,589.58)(0.785,0.496){43}{\rule{0.726pt}{0.120pt}}
\multiput(1152.00,588.17)(34.493,23.000){2}{\rule{0.363pt}{0.400pt}}
\multiput(1188.00,612.58)(0.949,0.495){31}{\rule{0.853pt}{0.119pt}}
\multiput(1188.00,611.17)(30.230,17.000){2}{\rule{0.426pt}{0.400pt}}
\multiput(1220.00,629.58)(1.091,0.493){23}{\rule{0.962pt}{0.119pt}}
\multiput(1220.00,628.17)(26.004,13.000){2}{\rule{0.481pt}{0.400pt}}
\multiput(1248.00,642.58)(1.225,0.491){17}{\rule{1.060pt}{0.118pt}}
\multiput(1248.00,641.17)(21.800,10.000){2}{\rule{0.530pt}{0.400pt}}
\multiput(1272.00,652.59)(1.484,0.488){13}{\rule{1.250pt}{0.117pt}}
\multiput(1272.00,651.17)(20.406,8.000){2}{\rule{0.625pt}{0.400pt}}
\multiput(1295.00,660.59)(1.756,0.482){9}{\rule{1.433pt}{0.116pt}}
\multiput(1295.00,659.17)(17.025,6.000){2}{\rule{0.717pt}{0.400pt}}
\put(516,534){\circle{12}}
\put(379,183){\circle{12}}
\put(538,217){\circle{12}}
\put(631,260){\circle{12}}
\put(697,291){\circle{12}}
\put(748,320){\circle{12}}
\put(790,342){\circle{12}}
\put(826,365){\circle{12}}
\put(856,383){\circle{12}}
\put(883,404){\circle{12}}
\put(985,471){\circle{12}}
\put(1055,523){\circle{12}}
\put(1108,560){\circle{12}}
\put(1152,589){\circle{12}}
\put(1188,612){\circle{12}}
\put(1220,629){\circle{12}}
\put(1248,642){\circle{12}}
\put(1272,652){\circle{12}}
\put(1295,660){\circle{12}}
\put(1315,666){\circle{12}}
\end{picture}

\end{center}

\centerline{ Figure  1 }
\end{figure}

\newpage

\begin{figure}
\begin{center}

\setlength{\unitlength}{0.240900pt}
\ifx\plotpoint\undefined\newsavebox{\plotpoint}\fi
\begin{picture}(1500,900)(0,0)
\font\gnuplot=cmr10 at 10pt
\gnuplot
\sbox{\plotpoint}{\rule[-0.200pt]{0.400pt}{0.400pt}}%
\put(220.0,113.0){\rule[-0.200pt]{4.818pt}{0.400pt}}
\put(198,113){\makebox(0,0)[r]{0.1}}
\put(1416.0,113.0){\rule[-0.200pt]{4.818pt}{0.400pt}}
\put(220.0,544.0){\rule[-0.200pt]{4.818pt}{0.400pt}}
\put(198,544){\makebox(0,0)[r]{0.3}}
\put(1416.0,544.0){\rule[-0.200pt]{4.818pt}{0.400pt}}
\put(220.0,745.0){\rule[-0.200pt]{4.818pt}{0.400pt}}
\put(198,745){\makebox(0,0)[r]{0.5}}
\put(1416.0,745.0){\rule[-0.200pt]{4.818pt}{0.400pt}}
\put(220.0,877.0){\rule[-0.200pt]{4.818pt}{0.400pt}}
\put(198,877){\makebox(0,0)[r]{0.7}}
\put(1416.0,877.0){\rule[-0.200pt]{4.818pt}{0.400pt}}
\put(220.0,113.0){\rule[-0.200pt]{0.400pt}{4.818pt}}
\put(220,68){\makebox(0,0){1}}
\put(220.0,857.0){\rule[-0.200pt]{0.400pt}{4.818pt}}
\put(394.0,113.0){\rule[-0.200pt]{0.400pt}{4.818pt}}
\put(394,68){\makebox(0,0){1.5}}
\put(394.0,857.0){\rule[-0.200pt]{0.400pt}{4.818pt}}
\put(567.0,113.0){\rule[-0.200pt]{0.400pt}{4.818pt}}
\put(567,68){\makebox(0,0){2}}
\put(567.0,857.0){\rule[-0.200pt]{0.400pt}{4.818pt}}
\put(741.0,113.0){\rule[-0.200pt]{0.400pt}{4.818pt}}
\put(741,68){\makebox(0,0){2.5}}
\put(741.0,857.0){\rule[-0.200pt]{0.400pt}{4.818pt}}
\put(915.0,113.0){\rule[-0.200pt]{0.400pt}{4.818pt}}
\put(915,68){\makebox(0,0){3}}
\put(915.0,857.0){\rule[-0.200pt]{0.400pt}{4.818pt}}
\put(1089.0,113.0){\rule[-0.200pt]{0.400pt}{4.818pt}}
\put(1089,68){\makebox(0,0){3.5}}
\put(1089.0,857.0){\rule[-0.200pt]{0.400pt}{4.818pt}}
\put(1262.0,113.0){\rule[-0.200pt]{0.400pt}{4.818pt}}
\put(1262,68){\makebox(0,0){4}}
\put(1262.0,857.0){\rule[-0.200pt]{0.400pt}{4.818pt}}
\put(1436.0,113.0){\rule[-0.200pt]{0.400pt}{4.818pt}}
\put(1436,68){\makebox(0,0){4.5}}
\put(1436.0,857.0){\rule[-0.200pt]{0.400pt}{4.818pt}}
\put(220.0,113.0){\rule[-0.200pt]{292.934pt}{0.400pt}}
\put(1436.0,113.0){\rule[-0.200pt]{0.400pt}{184.048pt}}
\put(220.0,877.0){\rule[-0.200pt]{292.934pt}{0.400pt}}
\put(45,495){\makebox(0,0){$R_L$}}
\put(828,23){\makebox(0,0){$1/T$}}
\put(220.0,113.0){\rule[-0.200pt]{0.400pt}{184.048pt}}
\put(1262,221){\raisebox{-.8pt}{\makebox(0,0){$\Box$}}}
\put(1031,317){\raisebox{-.8pt}{\makebox(0,0){$\Box$}}}
\put(865,393){\raisebox{-.8pt}{\makebox(0,0){$\Box$}}}
\put(741,461){\raisebox{-.8pt}{\makebox(0,0){$\Box$}}}
\put(567,540){\raisebox{-.8pt}{\makebox(0,0){$\Box$}}}
\put(452,617){\raisebox{-.8pt}{\makebox(0,0){$\Box$}}}
\put(369,665){\raisebox{-.8pt}{\makebox(0,0){$\Box$}}}
\put(1262,207){\usebox{\plotpoint}}
\multiput(1258.25,207.58)(-1.006,0.499){227}{\rule{0.903pt}{0.120pt}}
\multiput(1260.12,206.17)(-229.125,115.000){2}{\rule{0.452pt}{0.400pt}}
\multiput(1027.22,322.58)(-1.014,0.499){161}{\rule{0.910pt}{0.120pt}}
\multiput(1029.11,321.17)(-164.112,82.000){2}{\rule{0.455pt}{0.400pt}}
\multiput(861.26,404.58)(-1.002,0.499){121}{\rule{0.900pt}{0.120pt}}
\multiput(863.13,403.17)(-122.132,62.000){2}{\rule{0.450pt}{0.400pt}}
\multiput(737.26,466.58)(-1.002,0.499){171}{\rule{0.900pt}{0.120pt}}
\multiput(739.13,465.17)(-172.132,87.000){2}{\rule{0.450pt}{0.400pt}}
\multiput(563.23,553.58)(-1.011,0.499){111}{\rule{0.907pt}{0.120pt}}
\multiput(565.12,552.17)(-113.117,57.000){2}{\rule{0.454pt}{0.400pt}}
\multiput(448.22,610.58)(-1.016,0.498){79}{\rule{0.910pt}{0.120pt}}
\multiput(450.11,609.17)(-81.112,41.000){2}{\rule{0.455pt}{0.400pt}}
\end{picture}

\end{center}

\centerline{Figure 2}
\end{figure}

\newpage
\begin{figure}
\begin{center}
\setlength{\unitlength}{0.240900pt}
\ifx\plotpoint\undefined\newsavebox{\plotpoint}\fi
\begin{picture}(1500,900)(0,0)
\font\gnuplot=cmr10 at 10pt
\gnuplot
\sbox{\plotpoint}{\rule[-0.200pt]{0.400pt}{0.400pt}}%
\put(220.0,113.0){\rule[-0.200pt]{2.409pt}{0.400pt}}
\put(1426.0,113.0){\rule[-0.200pt]{2.409pt}{0.400pt}}
\put(220.0,149.0){\rule[-0.200pt]{2.409pt}{0.400pt}}
\put(1426.0,149.0){\rule[-0.200pt]{2.409pt}{0.400pt}}
\put(220.0,180.0){\rule[-0.200pt]{4.818pt}{0.400pt}}
\put(198,180){\makebox(0,0)[r]{1}}
\put(1416.0,180.0){\rule[-0.200pt]{4.818pt}{0.400pt}}
\put(220.0,390.0){\rule[-0.200pt]{2.409pt}{0.400pt}}
\put(1426.0,390.0){\rule[-0.200pt]{2.409pt}{0.400pt}}
\put(220.0,513.0){\rule[-0.200pt]{2.409pt}{0.400pt}}
\put(1426.0,513.0){\rule[-0.200pt]{2.409pt}{0.400pt}}
\put(220.0,600.0){\rule[-0.200pt]{2.409pt}{0.400pt}}
\put(1426.0,600.0){\rule[-0.200pt]{2.409pt}{0.400pt}}
\put(220.0,667.0){\rule[-0.200pt]{2.409pt}{0.400pt}}
\put(1426.0,667.0){\rule[-0.200pt]{2.409pt}{0.400pt}}
\put(220.0,722.0){\rule[-0.200pt]{2.409pt}{0.400pt}}
\put(1426.0,722.0){\rule[-0.200pt]{2.409pt}{0.400pt}}
\put(220.0,769.0){\rule[-0.200pt]{2.409pt}{0.400pt}}
\put(1426.0,769.0){\rule[-0.200pt]{2.409pt}{0.400pt}}
\put(220.0,810.0){\rule[-0.200pt]{2.409pt}{0.400pt}}
\put(1426.0,810.0){\rule[-0.200pt]{2.409pt}{0.400pt}}
\put(220.0,845.0){\rule[-0.200pt]{2.409pt}{0.400pt}}
\put(1426.0,845.0){\rule[-0.200pt]{2.409pt}{0.400pt}}
\put(220.0,877.0){\rule[-0.200pt]{4.818pt}{0.400pt}}
\put(198,877){\makebox(0,0)[r]{10}}
\put(1416.0,877.0){\rule[-0.200pt]{4.818pt}{0.400pt}}
\put(220.0,113.0){\rule[-0.200pt]{0.400pt}{4.818pt}}
\put(220,68){\makebox(0,0){0.01}}
\put(220.0,857.0){\rule[-0.200pt]{0.400pt}{4.818pt}}
\put(331.0,113.0){\rule[-0.200pt]{0.400pt}{2.409pt}}
\put(331.0,867.0){\rule[-0.200pt]{0.400pt}{2.409pt}}
\put(396.0,113.0){\rule[-0.200pt]{0.400pt}{2.409pt}}
\put(396.0,867.0){\rule[-0.200pt]{0.400pt}{2.409pt}}
\put(442.0,113.0){\rule[-0.200pt]{0.400pt}{2.409pt}}
\put(442.0,867.0){\rule[-0.200pt]{0.400pt}{2.409pt}}
\put(477.0,113.0){\rule[-0.200pt]{0.400pt}{2.409pt}}
\put(477.0,867.0){\rule[-0.200pt]{0.400pt}{2.409pt}}
\put(507.0,113.0){\rule[-0.200pt]{0.400pt}{2.409pt}}
\put(507.0,867.0){\rule[-0.200pt]{0.400pt}{2.409pt}}
\put(531.0,113.0){\rule[-0.200pt]{0.400pt}{2.409pt}}
\put(531.0,867.0){\rule[-0.200pt]{0.400pt}{2.409pt}}
\put(553.0,113.0){\rule[-0.200pt]{0.400pt}{2.409pt}}
\put(553.0,867.0){\rule[-0.200pt]{0.400pt}{2.409pt}}
\put(572.0,113.0){\rule[-0.200pt]{0.400pt}{2.409pt}}
\put(572.0,867.0){\rule[-0.200pt]{0.400pt}{2.409pt}}
\put(588.0,113.0){\rule[-0.200pt]{0.400pt}{4.818pt}}
\put(588,68){\makebox(0,0){0.1}}
\put(588.0,857.0){\rule[-0.200pt]{0.400pt}{4.818pt}}
\put(699.0,113.0){\rule[-0.200pt]{0.400pt}{2.409pt}}
\put(699.0,867.0){\rule[-0.200pt]{0.400pt}{2.409pt}}
\put(764.0,113.0){\rule[-0.200pt]{0.400pt}{2.409pt}}
\put(764.0,867.0){\rule[-0.200pt]{0.400pt}{2.409pt}}
\put(810.0,113.0){\rule[-0.200pt]{0.400pt}{2.409pt}}
\put(810.0,867.0){\rule[-0.200pt]{0.400pt}{2.409pt}}
\put(846.0,113.0){\rule[-0.200pt]{0.400pt}{2.409pt}}
\put(846.0,867.0){\rule[-0.200pt]{0.400pt}{2.409pt}}
\put(875.0,113.0){\rule[-0.200pt]{0.400pt}{2.409pt}}
\put(875.0,867.0){\rule[-0.200pt]{0.400pt}{2.409pt}}
\put(900.0,113.0){\rule[-0.200pt]{0.400pt}{2.409pt}}
\put(900.0,867.0){\rule[-0.200pt]{0.400pt}{2.409pt}}
\put(921.0,113.0){\rule[-0.200pt]{0.400pt}{2.409pt}}
\put(921.0,867.0){\rule[-0.200pt]{0.400pt}{2.409pt}}
\put(940.0,113.0){\rule[-0.200pt]{0.400pt}{2.409pt}}
\put(940.0,867.0){\rule[-0.200pt]{0.400pt}{2.409pt}}
\put(957.0,113.0){\rule[-0.200pt]{0.400pt}{4.818pt}}
\put(957,68){\makebox(0,0){1}}
\put(957.0,857.0){\rule[-0.200pt]{0.400pt}{4.818pt}}
\put(1068.0,113.0){\rule[-0.200pt]{0.400pt}{2.409pt}}
\put(1068.0,867.0){\rule[-0.200pt]{0.400pt}{2.409pt}}
\put(1132.0,113.0){\rule[-0.200pt]{0.400pt}{2.409pt}}
\put(1132.0,867.0){\rule[-0.200pt]{0.400pt}{2.409pt}}
\put(1179.0,113.0){\rule[-0.200pt]{0.400pt}{2.409pt}}
\put(1179.0,867.0){\rule[-0.200pt]{0.400pt}{2.409pt}}
\put(1214.0,113.0){\rule[-0.200pt]{0.400pt}{2.409pt}}
\put(1214.0,867.0){\rule[-0.200pt]{0.400pt}{2.409pt}}
\put(1243.0,113.0){\rule[-0.200pt]{0.400pt}{2.409pt}}
\put(1243.0,867.0){\rule[-0.200pt]{0.400pt}{2.409pt}}
\put(1268.0,113.0){\rule[-0.200pt]{0.400pt}{2.409pt}}
\put(1268.0,867.0){\rule[-0.200pt]{0.400pt}{2.409pt}}
\put(1289.0,113.0){\rule[-0.200pt]{0.400pt}{2.409pt}}
\put(1289.0,867.0){\rule[-0.200pt]{0.400pt}{2.409pt}}
\put(1308.0,113.0){\rule[-0.200pt]{0.400pt}{2.409pt}}
\put(1308.0,867.0){\rule[-0.200pt]{0.400pt}{2.409pt}}
\put(1325.0,113.0){\rule[-0.200pt]{0.400pt}{4.818pt}}
\put(1325,68){\makebox(0,0){10}}
\put(1325.0,857.0){\rule[-0.200pt]{0.400pt}{4.818pt}}
\put(1436.0,113.0){\rule[-0.200pt]{0.400pt}{2.409pt}}
\put(1436.0,867.0){\rule[-0.200pt]{0.400pt}{2.409pt}}
\put(220.0,113.0){\rule[-0.200pt]{292.934pt}{0.400pt}}
\put(1436.0,113.0){\rule[-0.200pt]{0.400pt}{184.048pt}}
\put(220.0,877.0){\rule[-0.200pt]{292.934pt}{0.400pt}}
\put(45,495){\makebox(0,0){$E/J R_L$}}
\put(828,23){\makebox(0,0){$J/T^{1+\nu}$}}
\put(220.0,113.0){\rule[-0.200pt]{0.400pt}{184.048pt}}
\put(477,769){\makebox(0,0)[r]{T=0.6}}
\put(521,769){\raisebox{-.8pt}{\makebox(0,0){$\Diamond$}}}
\put(486,180){\raisebox{-.8pt}{\makebox(0,0){$\Diamond$}}}
\put(597,181){\raisebox{-.8pt}{\makebox(0,0){$\Diamond$}}}
\put(662,186){\raisebox{-.8pt}{\makebox(0,0){$\Diamond$}}}
\put(708,191){\raisebox{-.8pt}{\makebox(0,0){$\Diamond$}}}
\put(744,199){\raisebox{-.8pt}{\makebox(0,0){$\Diamond$}}}
\put(773,204){\raisebox{-.8pt}{\makebox(0,0){$\Diamond$}}}
\put(797,211){\raisebox{-.8pt}{\makebox(0,0){$\Diamond$}}}
\put(819,220){\raisebox{-.8pt}{\makebox(0,0){$\Diamond$}}}
\put(838,228){\raisebox{-.8pt}{\makebox(0,0){$\Diamond$}}}
\put(908,264){\raisebox{-.8pt}{\makebox(0,0){$\Diamond$}}}
\put(957,300){\raisebox{-.8pt}{\makebox(0,0){$\Diamond$}}}
\put(995,331){\raisebox{-.8pt}{\makebox(0,0){$\Diamond$}}}
\put(1025,358){\raisebox{-.8pt}{\makebox(0,0){$\Diamond$}}}
\put(1050,381){\raisebox{-.8pt}{\makebox(0,0){$\Diamond$}}}
\put(1072,400){\raisebox{-.8pt}{\makebox(0,0){$\Diamond$}}}
\put(477,724){\makebox(0,0)[r]{T=0.5}}
\put(521,724){\makebox(0,0){$+$}}
\put(542,180){\makebox(0,0){$+$}}
\put(652,188){\makebox(0,0){$+$}}
\put(717,196){\makebox(0,0){$+$}}
\put(763,207){\makebox(0,0){$+$}}
\put(799,215){\makebox(0,0){$+$}}
\put(828,225){\makebox(0,0){$+$}}
\put(853,238){\makebox(0,0){$+$}}
\put(874,247){\makebox(0,0){$+$}}
\put(893,259){\makebox(0,0){$+$}}
\put(964,306){\makebox(0,0){$+$}}
\put(1013,347){\makebox(0,0){$+$}}
\put(1050,384){\makebox(0,0){$+$}}
\put(1080,414){\makebox(0,0){$+$}}
\put(1106,439){\makebox(0,0){$+$}}
\put(1128,459){\makebox(0,0){$+$}}
\put(477,679){\makebox(0,0)[r]{T=0.4}}
\put(521,679){\raisebox{-.8pt}{\makebox(0,0){$\Box$}}}
\put(609,180){\raisebox{-.8pt}{\makebox(0,0){$\Box$}}}
\put(720,184){\raisebox{-.8pt}{\makebox(0,0){$\Box$}}}
\put(785,198){\raisebox{-.8pt}{\makebox(0,0){$\Box$}}}
\put(831,214){\raisebox{-.8pt}{\makebox(0,0){$\Box$}}}
\put(867,232){\raisebox{-.8pt}{\makebox(0,0){$\Box$}}}
\put(896,248){\raisebox{-.8pt}{\makebox(0,0){$\Box$}}}
\put(921,263){\raisebox{-.8pt}{\makebox(0,0){$\Box$}}}
\put(942,274){\raisebox{-.8pt}{\makebox(0,0){$\Box$}}}
\put(961,290){\raisebox{-.8pt}{\makebox(0,0){$\Box$}}}
\put(1032,350){\raisebox{-.8pt}{\makebox(0,0){$\Box$}}}
\put(1080,400){\raisebox{-.8pt}{\makebox(0,0){$\Box$}}}
\put(1118,440){\raisebox{-.8pt}{\makebox(0,0){$\Box$}}}
\put(1148,473){\raisebox{-.8pt}{\makebox(0,0){$\Box$}}}
\put(1174,499){\raisebox{-.8pt}{\makebox(0,0){$\Box$}}}
\put(1196,521){\raisebox{-.8pt}{\makebox(0,0){$\Box$}}}
\put(477,634){\makebox(0,0)[r]{T=0.35}}
\put(521,634){\makebox(0,0){$\times$}}
\put(650,180){\makebox(0,0){$\times$}}
\put(761,192){\makebox(0,0){$\times$}}
\put(826,212){\makebox(0,0){$\times$}}
\put(872,238){\makebox(0,0){$\times$}}
\put(907,258){\makebox(0,0){$\times$}}
\put(937,278){\makebox(0,0){$\times$}}
\put(961,296){\makebox(0,0){$\times$}}
\put(983,313){\makebox(0,0){$\times$}}
\put(1002,328){\makebox(0,0){$\times$}}
\put(1072,397){\makebox(0,0){$\times$}}
\put(1121,451){\makebox(0,0){$\times$}}
\put(1158,492){\makebox(0,0){$\times$}}
\put(1189,526){\makebox(0,0){$\times$}}
\put(1214,553){\makebox(0,0){$\times$}}
\put(1236,574){\makebox(0,0){$\times$}}
\put(477,589){\makebox(0,0)[r]{T=0.3}}
\put(521,589){\makebox(0,0){$\triangle$}}
\put(697,180){\makebox(0,0){$\triangle$}}
\put(808,198){\makebox(0,0){$\triangle$}}
\put(873,233){\makebox(0,0){$\triangle$}}
\put(919,263){\makebox(0,0){$\triangle$}}
\put(954,290){\makebox(0,0){$\triangle$}}
\put(984,318){\makebox(0,0){$\triangle$}}
\put(1008,336){\makebox(0,0){$\triangle$}}
\put(1030,356){\makebox(0,0){$\triangle$}}
\put(1048,375){\makebox(0,0){$\triangle$}}
\put(1119,451){\makebox(0,0){$\triangle$}}
\put(1168,508){\makebox(0,0){$\triangle$}}
\put(1205,552){\makebox(0,0){$\triangle$}}
\put(1236,586){\makebox(0,0){$\triangle$}}
\put(1261,613){\makebox(0,0){$\triangle$}}
\put(1283,634){\makebox(0,0){$\triangle$}}
\put(477,544){\makebox(0,0)[r]{T=0.25}}
\put(521,544){\makebox(0,0){$\star$}}
\put(752,180){\makebox(0,0){$\star$}}
\put(863,220){\makebox(0,0){$\star$}}
\put(928,271){\makebox(0,0){$\star$}}
\put(974,309){\makebox(0,0){$\star$}}
\put(1010,343){\makebox(0,0){$\star$}}
\put(1039,369){\makebox(0,0){$\star$}}
\put(1064,396){\makebox(0,0){$\star$}}
\put(1085,418){\makebox(0,0){$\star$}}
\put(1104,443){\makebox(0,0){$\star$}}
\put(1174,522){\makebox(0,0){$\star$}}
\put(1223,583){\makebox(0,0){$\star$}}
\put(1261,628){\makebox(0,0){$\star$}}
\put(1291,662){\makebox(0,0){$\star$}}
\put(1316,689){\makebox(0,0){$\star$}}
\put(1338,709){\makebox(0,0){$\star$}}
\end{picture}

\end{center}

\centerline{Figure 3}
\end{figure}

\end{document}